\documentclass[10pt]{article}
\usepackage{amsmath}
\usepackage{graphicx}
\usepackage{siunitx}
\usepackage[font={footnotesize}, labelfont={bf}]{caption}
\usepackage[left=1.5in, right=1.5in]{geometry}
\usepackage{lineno}
\usepackage[final]{changes}

\usepackage{fancyhdr}
\usepackage[
    backend=biber,
    style=authoryear-comp,
    maxcitenames=99,
    maxbibnames=99,
    uniquename=init,
    giveninits,
    uniquelist=false,
    sortcites,
    sorting=nyt
]{biblatex}
\renewbibmacro{in:}{}
\DeclareNameAlias{sortname}{family-given}

\usepackage{setspace}
\title{Critical Analysis of Mediterranean Sea Level Limit Cycles During the Messinian Salinity Crisis}
\author{\textsc{Mark M. Baum}$^{1,}$\thanks{Corresponding Author: \texttt{markbaum@g.harvard.edu}}}
\date{\today}

\begin{document}

\maketitle
\thispagestyle{empty}

\vspace{-0.8cm}
\begin{center}
    \footnotesize
    $^1$Department of Earth and Planetary Sciences, Harvard University, Cambridge MA, 02138, USA
\end{center}

\begin{center}
\end{center}
\vspace{0.5cm}

\begin{abstract}
\singlespacing
The Messinian Salinity Crisis (5.97-5.33~Ma) may be one of the most significant periods of sea-level change in recent geologic history. During this period, evaporite deposition throughout the Mediterranean basin records a series of dramatic environmental changes as flow through the Strait of Gibraltar was restricted. In the first stage of evaporite deposition, cycles of gypsum appear in shallow basins on the margins of the Mediterranean. The complex environmental history giving rise to these cycles has been investigated for decades but remains controversial. Notably, whether the evaporites are connected to significant changes in Mediterranean sea-level is an open question. In one proposed model, competition between tectonic uplift and erosion at the Strait of Gibraltar gives rise to self-sustaining sea-level oscillations---limit cycles---which trigger evaporite deposition. Here I show that limit cycles are not a robust result of the proposed model and discuss how any oscillations produced by this model depend on an unrealistic formulation of a key model equation. First, I simplify the model equations and test whether limit cycles are produced in 64 million unique combinations of model parameters, finding oscillations in only 0.2~\% of all simulations. Next, I examine the formulation of a critical model equation representing stream channel slope over the Strait of Gibraltar, concluding that a more realistic formulation would render sea-level limit cycles improbable, if not impossible, in the proposed model.
\end{abstract}

\newpage
\section*{INTRODUCTION}

At the end of the Miocene, a combination of environmental and tectonic processes dramatically altered the water budget and chemistry of the Mediterranean Sea, leading to the Messinian Salinity Crisis (MSC) between 5.97 and 5.33~Ma \parencite{manzi_age_2013}. This period is marked in the stratigraphic record by evaporite deposition throughout the basin \parencite{hsu_late_1973}. It is generally thought that the crisis unfolded in three stages \parencite{briand_messinian_2008, roveri_messinian_2014}. In the first stage, cycles of gypsum and marine marl were deposited in shallow, marginal basins. In stage two, massive halite deposits formed in deeper basins. In the third and final ``Lagomare" stage, large fluctuations in salinity are exhibited by gypsum-marl pairs and evaporite-free deposits.

However, fundamental questions about the causes and timing of events during the MSC remain. A precise geologic and environmental chronology is elusive because of the paucity of marine fossils in Messinian sediments and lithologies in the basin that are not well suited for accurate dating \parencite{roveri_messinian_2014}. The first stage, recorded by 16-17 cycles of gypsum deposition separated by shale and carbonate \parencite{krijgsman_chronology_1999, rohling_controls_2008, lugli_primary_2010}, is particularly enigmatic. Various mechanisms have been proposed to explain the cycles, generally focusing on some combination of orbitally forced climate change, variation in Antarctic ice mass, and the dynamics of tectonic uplift and erosion at the Strait of Gibraltar \parencite{krijgsman_chronology_1999, krijgsman_astrochronology_2001, gargani_mediterranean_2007, lugli_primary_2010, garcia-castellanos_messinian_2011, roveri_messinian_2014, ohneiser_antarctic_2015, simon_salinity_2017}. There is, however, no consensus on the cause of evaporite cycles in the first stage of the MSC \parencite{de_lange_messinian_2010, lugli_primary_2010}. They are an intriguing mystery at the intersection of environmental and geological history.

In one explanation of the cycles, \textcite{garcia-castellanos_messinian_2011} proposed an elegant model where competition between tectonic uplift and erosion at the Strait of Gibraltar gives rise to oscillation of Mediterranean sea-level (MSL), repeatedly triggering gypsum deposition. Their model exhibits limit cycles \parencite{strogatz_nonlinear_1994}, where no oscillatory external forcing, such as Milankovitch cycles, is required to produce oscillations in Mediterranean sea level. If correct, this model would constitute an unusual and fascinating example of large-scale sea-level change that is not driven by redistribution of water between the oceans and the cryosphere.

There are geological reasons to doubt this model. \added{While a major erosional surface is found at the top of the unit containing the gypsum beds, no subaerial erosional surfaces have been observed at the tops of individual gypsum beds and they are thought to have formed in waters shallower than 200 m} \parencite{lugli_primary_2010}. \replaced{The model, however,}{It} exhibits $>$400~m fluctuations in Mediterranean sea-level, \replaced{considerably greater than the observations indicate.}{but no subaerial erosional surfaces have been discovered at the top of gypsum beds and they are thought to have formed in waters shallower than 200~m.} 

Here I show that there are also computational and conceptual reasons to doubt the model. In the Model Formulation section, I consolidate the original model equations into a system of two explicit, analytic, ordinary differential equations (ODEs). In the Simulations section, I describe how these equations are solved, explain my computational approach to identifying limit cycles, and show how likely limit cycles are to occur in different parameter ranges. In the Revisiting Channel Slope section, I examine a key component of the model---the channel slope---and how it relates to the model's capacity to produce oscillations. Finally, in the Discussion \& Conclusion section, I review the implications of the prior sections, summarize my conclusions, and briefly describe potential future work.

\section*{MODEL FORMULATION}

\subsection*{Original Model}

The defining characteristic of the MSL limit cycle model \parencite{garcia-castellanos_messinian_2011} is its capacity to produce oscillations without any external periodic forcing. Water is exchanged solely between the Mediterranean Sea and the ocean, without any forcing by the background climate or polar ice mass. The model includes three primary physical processes. First, the height of the sill at Gibraltar, where water flows from the Atlantic to the Mediterranean, is controlled by fluvial erosion and tectonic uplift. Second, sea level in the Mediterranean responds to a water budget including discharge over the sill, evaporation, direct precipitation, and input from continental rivers. Third, sea level in the ocean, outside the sill, changes to conserve the water lost and received by the Mediterranean.

As the original authors describe, the competition between uplift and erosion at the sill appears to give rise to an oscillatory coupling between erosion and Mediterranean sea level, possibly explaining the cyclic evaporite deposits in the first stage of the MSC. In this conception, the Strait of Gibraltar is initially open and the Mediterranean is full. As the sill is slowly uplifted, the sill depth is reduced and flow to the Mediterranean is restricted, causing MSL to drop. Next, with the equations used, the increased hydraulic head difference between the Atlantic and Mediterranean causes a nonlinearly accelerating increase in erosion at the sill. Increased erosion deepens the sill, enabling greater flow to the Mediterranean, which refills the basin and raises MSL. Then uplift continues and the cycle repeats.

More recently, \textcite{coulson_role_2019} illustrated the importance of additional sea-level physics as the Strait of Gibraltar opens and closes, extending the original model. They coupled a sophisticated sea-level model to the original erosion-uplift equations \parencite{garcia-castellanos_messinian_2011}, incorporating the effects of self-gravitation in the water bodies and crustal deformation in response to changing water load. With the additional physics, the model still exhibits limit cycles, but with a slower uplift rate at Gibraltar that is more consistent with estimates from independent geodynamical models \parencite{duggen_deep_2003, gerya_thermomechanical_2004, andrews_rheologic_2009, duretz_numerical_2011}. The result appears to strengthen the idea of limit cycles during the first stage of the MSC, as the inclusion of well-established sea-level theory brings a key model parameter, the uplift rate, closer to expected values.

\subsection*{Simplified Model Equations}

The original model has four dependent variables: ocean sea-level, Gibraltar sill height, western Mediterranean sea level, and eastern Mediterranean sea level \parencite{garcia-castellanos_messinian_2011}. First, to consolidate and simplify the system, I assume that the Mediterranean behaves as a single basin, setting aside the separate treatment of the western and eastern regions. The Sicily sill, which separates the eastern and western basins, may have played a role when MSL was low \parencite{just_erosion_2011}. However, in the original model, oscillations are shown to occur almost entirely above the Sicily sill level (-430~m), so the exclusion of the sill should not alter the dynamics of the entire system. This assumption will be further discussed in the final section.

\begin{figure}[ht]
	\centering
	\includegraphics[width=0.8\textwidth]{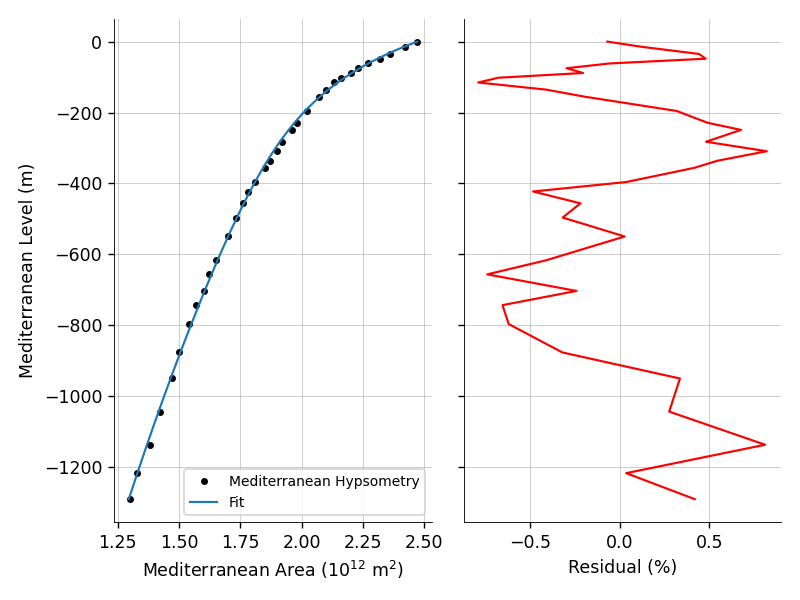}
	\caption{The area of the Mediterranean Sea as a function of sea level (from \parencite{meijer_quantitative_2005}), the fit described by equation (\ref{eq:Am}), and the residual. The fit is accurate to better than 1~\% down to at least 1200 m.}
	\label{fig:Am}
\end{figure}

Next, I develop simple analytical expressions for two important components of the model, the surface area of the Mediterranean and the level of the ocean. The original model implemented linear interpolation of hypsometric curves for the Mediterranean surface area as a function of Mediterranean sea level during the Messinian \parencite{meijer_quantitative_2005}, using roughly 10 points along the curve. Instead, I fit a sum of two exponentials to the modern Mediterranean surface area curve. Modern hypsometry and modeled Messinian hypsometry are nearly the same in the upper 2.5~km of the basin \parencite{meijer_quantitative_2005}, so either one can reasonably be used. The sum of two exponentials representing the surface area of the Mediterranean is
\begin{linenomath*}
  \begin{equation}
    A_m(z_m) = c_1 e^{z_m/\alpha_1} + c_2 e^{z_m/\alpha_2} \, ,
    \label{eq:Am}
  \end{equation}
\end{linenomath*}
where $A_m$ is the Mediterranean surface area, $z_m$ is MSL, and parameters $c_1$, $c_2$, $\alpha_1$, and $\alpha_2$ and are found by fitting equation (\ref{eq:Am}) to present-day hypsometry in Figure 2 of \parencite{meijer_quantitative_2005}. The values of these parameters are shown in Table \ref{tab:Am}. Figure \ref{fig:Am} shows the result of the fit and the residual, which is less than 1~\% to a depth of at least 1200~m. Equation (\ref{eq:Am}) is simpler than linear interpolation and a slightly better representation of a smoothly varying surface area. It also has the advantage that it can be used to formulate an expression for the ocean sea level, as I explain next.

The model assumes all water lost from the Mediterranean is instantaneously received by the oceans, and vice versa, conserving water. This a reasonable assumption considering the timescales of evaporation and precipitation. Because of this assumption, the ocean level $z_o$ does not require a time-dependent model equation. It is solely dependent on the level of the Mediterranean. A tiny change in the Mediterranean level, $dz_m$, causes a tiny, opposing change in ocean level, $dz_o$.
\begin{linenomath*}
  \begin{equation}
  	A_o \, dz_o = -A_m \, dz_m
  	\label{eq:dzo}
  \end{equation}
\end{linenomath*}
Equation (\ref{eq:dzo}) is a statement of volume conservation. Because $A_o$ is so large (about $360\times10^{12}$~m$^2$), I let it be a constant. Even a 20~m change in $z_o$, much larger than the modeled range of values, would only cause $A_o$ to change by about 1~\% with the current hypsometry of world oceans \parencite{noauthor_volumes_2010, noauthor_hypsographic_2012}, so it is acceptable to use a fixed value for $A_o$.

Integrating equation (\ref{eq:dzo}), using equation (\ref{eq:Am}) for $A_m$, produces an expression for the ocean level as a function of the Mediterranean level,
\begin{linenomath*}
  \begin{equation}
  	z_o(z_m) = \frac{c_1 \alpha_1}{A_o} \left( 1 - e^{z_m/\alpha_1} \right) + \frac{c_2 \alpha_2}{A_o} \left( 1 - e^{z_m/\alpha_2} \right) \, .
  	\label{eq:zo}
  \end{equation}
\end{linenomath*}
This expression obviates the need for a time-dependent equation defining $z_o$. Because $z_o$ is a function of $z_m$, the only time-dependent variables in the system are the sill height and MSL.

Next, I combine the equations for erosion and discharge at the sill into two ODEs for the sill height $z_s$ and Mediterranean level $z_m$. These are now the only two time-dependent variables in the model. \textcite{garcia-castellanos_messinian_2011} model erosion at the sill with a simple power function of basal shear stress in the river channel connecting the Atlantic and Mediterranean,
\begin{linenomath*}
  \begin{equation}
  	\dot{z}_s = U - k_b \max \left[ \tau - \tau_c \, , \, 0 \right]^a \, ,
  	\label{eq:zsdot}
  \end{equation}
\end{linenomath*}
where $\dot{z}_s$ is the time derivative of the sill height ($dz_s/dt$), $U$ is the uplift rate at the sill, $k_b$ is an erodibility coefficient, $\tau$ is the shear stress exerted by the flowing water, $\tau_c$ is the critical shear stress, and $a$ is an erosion exponent. The ``max" operator prevents erosion from occurring when $\tau < \tau_c$. The shear stress $\tau$ is a function of the channel's depth and slope,
\begin{linenomath*}
  \begin{equation}
  	\tau = \rho	g (z_o - z_s) S \, ,
  	\label{eq:tau}
  \end{equation}
\end{linenomath*}
where $\rho$ is water's density, $g$ is gravitational acceleration, $z_o - z_s$ is the approximate depth of the channel, and $S$ is the slope of the water surface. \textcite{garcia-castellanos_messinian_2011} compute the slope using
\begin{linenomath*}
  \begin{equation}
  	S = \frac{z_o - z_m}{L} \, ,
  	\label{eq:S}
  \end{equation}
\end{linenomath*}
where $L$ is a constant length of 100~km, representing the approximate half-width of the Betic-Rifean orogen \parencite{garcia-castellanos_messinian_2011}. Equation (\ref{eq:S}) is meant to approximate the mean channel slope. Plugging equations (\ref{eq:S}) and (\ref{eq:tau}) into equation (\ref{eq:zsdot}) yields
\begin{linenomath*}
  \begin{equation}
  	\dot{z}_s = U - k_b \max \left[ \frac{\rho g}{L}(z_o - z_s)(z_o - z_m) - \tau_c  \, , \, 0 \right]^a \, ,
  \label{eq:zsdot2}
  \end{equation}
\end{linenomath*}
a single expression for the rate of change of the sill height that depends only on $z_s$ and $z_m$ because $z_o$ is a function of $z_m$ defined by equation (\ref{eq:zo}).

The second component of the model is the level of the Mediterranean. Changes in MSL are governed by input and removal of water,
\begin{linenomath*}
  \begin{equation}
  	\dot{z}_m = P - E + \frac{R + Q}{A_m} \, ,
  	\label{eq:zmdot}
  \end{equation}
\end{linenomath*}
where $P$ is direct precipitation into the Mediterranean, $E$ is evaporation, $R$ accounts for input from continental rivers, and $Q$ is input from the ocean. By design, discharge over the sill only occurs from the Atlantic into the Mediterranean, without any return flow. The Mediterranean area $A_m$ is computed with equation (\ref{eq:Am}). \textcite{garcia-castellanos_messinian_2011} compute $Q$ with a simple geometric relationship,
\begin{linenomath*}
  \begin{equation}
  	Q = W(z_o - z_s)V \, ,
  	\label{eq:Q}
  \end{equation}
\end{linenomath*}
where $W$ is the width of the channel flowing over the sill, $z_o - z_s$ is the channel depth, and $V$ is the flow velocity. The velocity is represented by Manning's formula,
\begin{linenomath*}
  \begin{equation}
  	V = \frac{1}{n}R_h^{2/3}S^{1/2} \, ,
  	\label{eq:V}
  \end{equation}
\end{linenomath*}
where $n$ is a roughness coefficient, $R_h$ is the hydraulic radius, and $S$ is again the slope. Because the channel depth is expected to be considerably smaller than the channel width, the hydraulic radius is approximated by the channel depth, $R_h = z_o - z_s$. \textcite{garcia-castellanos_messinian_2011} choose a form for the channel width $W$ that accounts for the effect of uplift \parencite{turowski_cover_2007},
\begin{linenomath*}
  \begin{equation}
  	W = C_w \left( \frac{\tau_c + U/k_b}{\rho g} \right)^{-3/13} \left( nQ \right)^{6/13} \, ,
  	\label{eq:W}
  \end{equation}
\end{linenomath*}
where $C_w$ is an empirical constant. For clarity, I let
\begin{linenomath*}
  \begin{equation}
	T \equiv C_w \left( \frac{\tau_c + U/k_b}{\rho g} \right)^{-3/13} \, .
	\label{eq:T}
  \end{equation}
\end{linenomath*}
By plugging equations (\ref{eq:W}) and (\ref{eq:V}) into equation (\ref{eq:Q}), then plugging the result into equation (\ref{eq:zmdot}), the expression for $\dot{z}_m$ becomes
\begin{linenomath*}
  \begin{equation}
	\dot{z}_m = P - E + \frac{1}{A_m}\left[ R + \frac{T^{13/7}}{n}\max[z_o - z_s\,,0]^{65/21} \left( \frac{z_o - z_m}{L} \right)^{13/14} \right] \, .
	\label{eq:zmdot2}
  \end{equation}
\end{linenomath*}
Here the ``max" operator handles cases where the sill becomes higher than the ocean. When this occurs, the operator yields zero and flow from the ocean is shut off, disconnecting the Mediterranean and allowing the sill to rise indefinitely. Equation (\ref{eq:zmdot2}) is a consolidated, explicit form for the rate of change of MSL in the model of \parencite{garcia-castellanos_messinian_2011}, treating the Mediterranean as a single basin.

In summary, equations (\ref{eq:Am}), (\ref{eq:zo}), (\ref{eq:zsdot2}), and (\ref{eq:zmdot2}) constitute a simplified and consolidated version of the original model from \textcite{garcia-castellanos_messinian_2011}, treating the Mediterranean as a single basin instead of splitting it into eastern and western basins separated by the Sicily sill at -430~m.
\begin{itemize}
    \singlespacing
    \item Equation (\ref{eq:Am}) defines the Mediterranean Sea's surface area as MSL varies, $A_m$
    \item Equation (\ref{eq:zo}) defines the level of the ocean outside the sill at the Strait of Gibraltar, $z_o$
    \item Equation (\ref{eq:zsdot2}) defines the time rate-of-change of the sill height, $\dot{z}_s$
    \item Equation (\ref{eq:zmdot2}) defines the time rate-of-change of Mediterranean sea-level, $\dot{z}_m$
\end{itemize}

These four equations represent a system of two ODEs with two supporting expressions and ten physical parameters. For a summary of parameters and their values, see Tables \ref{tab:Am} and \ref{tab:param}. I will refer to the four equations listed above simply as ``the model." All of the expressions are analytic, without the need for external data or interpolation. The system is readily solved by standard numerical ODE integration techniques.

\section*{SIMULATIONS}

To better understand the conditions that produce MSL oscillation with this model, I solve the equations with a range of parameter values. The model equations were implemented in Python and C++, then integrated using SciPy's LSODA wrapper \parencite{scipy} and Verner's ``most efficient" 6th order Runge-Kutta pair \parencite{verner_numerically_2010}, respectively. Both methods utilize adaptive step size selection so that integration proceeds extremely rapidly when the solution is stable. As a check on the code, results from the two implementations were compared for sets of identical input parameters. They were indistinguishable in all tested cases.

First, I integrate the model using the same parameter values as the main results of prior work \parencite{garcia-castellanos_messinian_2011}. Table \ref{tab:param} shows these parameter values in the ``reference value" column. The parameter $C_w$ is an empirically determined constant, set to six here as it appears to be in prior modeling. As in prior modeling, integration is carried out for 100~kyr and the initial conditions are $z_s = -60$~m and $z_m = 0$~m. Although it does not affect the results, the initial sill depth of -60~m is considerably less than the present depth of the Camarinal Sill, which is about 300~m.

\begin{figure}[ht]
	\centering
	\includegraphics[width=0.75\textwidth]{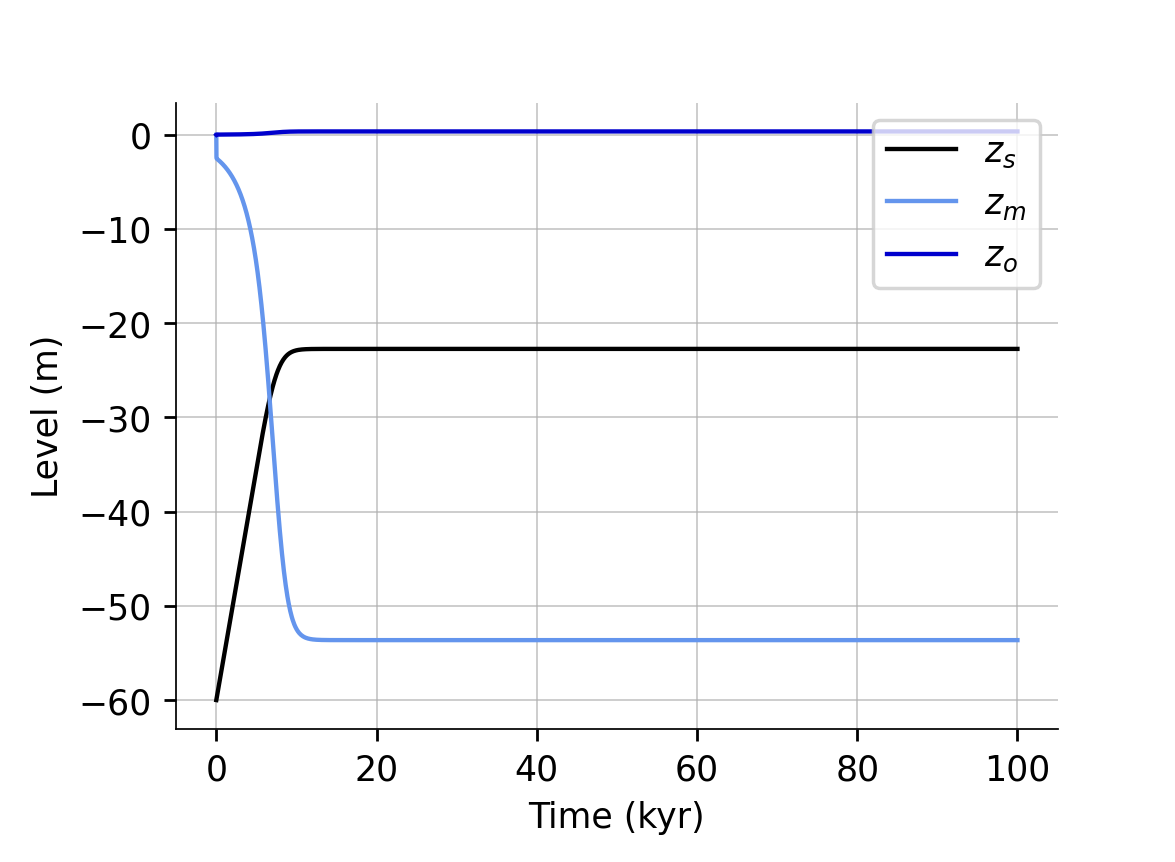}
	\caption{Results for the reference value parameters listed in Table \ref{tab:param}. As the legend indicates, the black line shows the sill level, the light blue line shows the Mediterranean level, and the dark blue line shows the ocean level.}
	\label{fig:refsim}
\end{figure}

Figure \ref{fig:refsim} shows results for the reference parameter values. First, MSL rapidly drops several meters over about 25 years. This is an initial adjustment to the Mediterranean's imbalanced water budget which starts the flow of water over the sill. After about 10~kyr, the model comes to rest with the Mediterranean level at roughly -54~m and the sill level at about -23~m. Uplift at the sill is exactly matched by erosion and the Mediterranean water budget is balanced. No oscillation is produced with the reference parameter values.

Next, I integrate the model for wide ranges of six key parameters. The ``tested range" column of Table \ref{tab:param} shows the minimum and maximum value tested for each parameter. A set of twenty values spanning each range was generated. The values are uniformly spaced for all parameters except $k_b$, where they are logarithmically spaced. Each of the ranges is roughly centered on the reference value except $U$ because the uplift rate is thought to be lower than the reference value of 4.9~mm/yr used in the original model \parencite{duggen_deep_2003, gerya_thermomechanical_2004, andrews_rheologic_2009, duretz_numerical_2011, coulson_role_2019}. For all possible combinations of these parameter ranges---a total of 20$^6$ or 64 million combinations---the model was integrated and checked for oscillation using the steps described next.

To determine whether a set of parameters produces oscillations, the model is integrated in 25 kyr intervals for a maximum duration of 100 Myr, beginning with the initial conditions $z_s = -60$~m and $z_m = 0$~m. After each 25 kyr integration interval, the state of the system is checked. If the sill is above the ocean ($z_s>z_o$), erosion and flow have ceased and the Mediterranean is disconnected. In this case, the sill will rise indefinitely, the system is not oscillating, and integration stops. Similarly, if the system appears to reach a stable fixed point, where $z_s$ and $z_m$ are no longer changing ($\dot{z}_s = 0$ and $\dot{z}_m = 0$), then the system is not oscillating and integration stops. If integration proceeds for the entire 100 Myr limit without Mediterranean disconnection or arrival at a fixed point, the solution is assumed to be oscillating.

To determine if the system is at (or very near) a stable fixed point, a two-stage check is performed.
\begin{enumerate}
\item The magnitudes of $\dot{z}_s$ and $\dot{z}_m$ are both less than 1 micrometer per year (3.169$\times 10^{-14}$ m/s)
\item A fixed point exists near the current model state and both system variables are within 0.01 \% of the fixed point's coordinates.
\end{enumerate}
If the first check is satisfied, the second check is performed. If the second check is also satisfied, the model is assumed to be stationary and integration stops. Fixed points are located using multivariate Newton's method and a finite difference approximation of the system's Jacobian. Performing this check in two stages prevents the initiation of Newton's method when the solution is not near a fixed point. This stringent, two-stage test is designed to prevent any oscillatory solutions from being erroneously identified as stable ones.

\begin{figure}[ht]
	\centering
	\includegraphics[width=\textwidth]{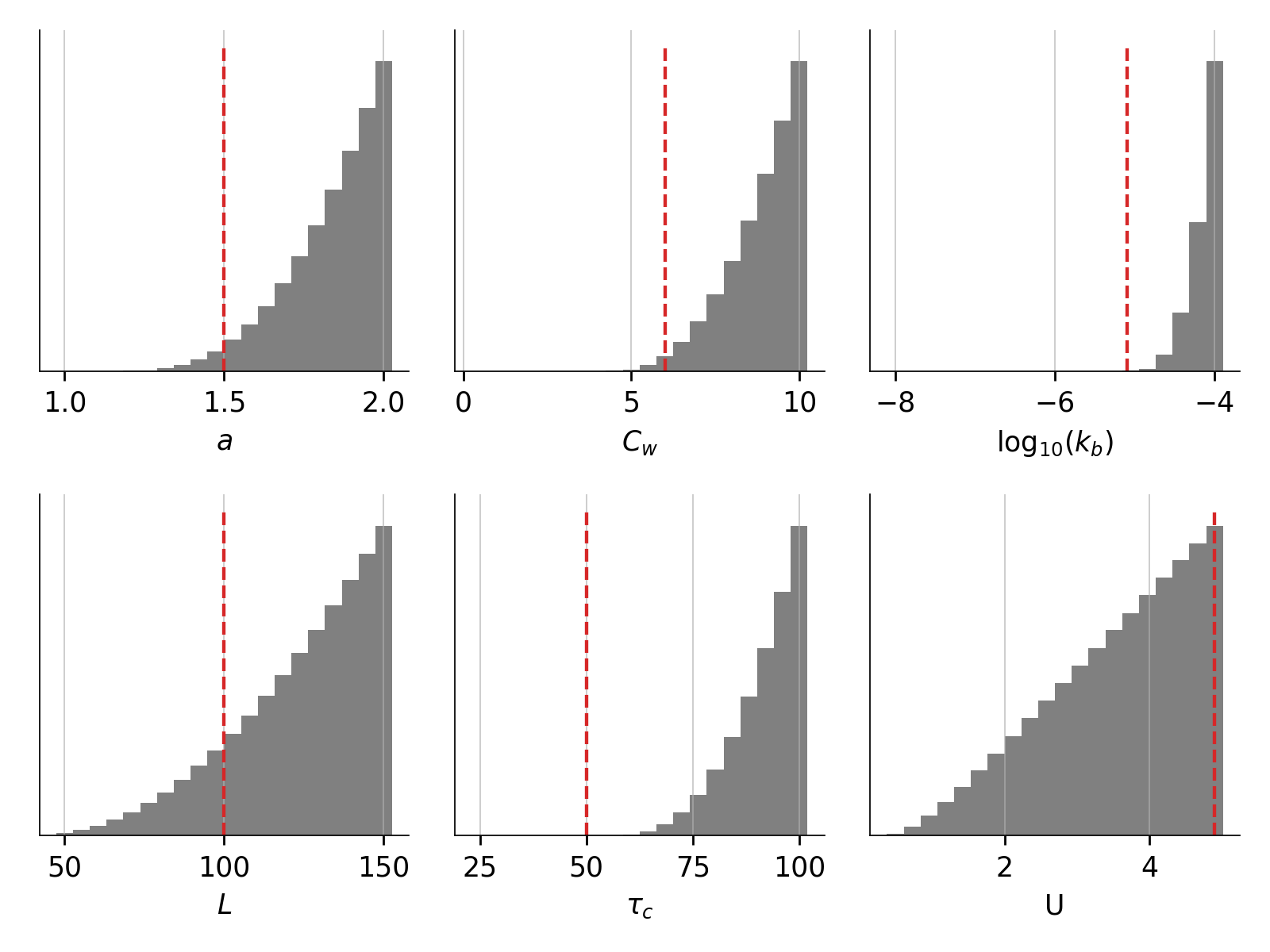}
	\caption{The distribution of oscillatory solutions for each parameter value. Each histogram shows all oscillatory solutions, arranged by the range of values for the given parameter. The total number of oscillatory solutions represents roughly 0.2~\% of all simulations. Red dashed lines indicate reference values (see Table \ref{tab:param}). Notably, there is a complete lack of oscillatory solutions near the reference values of $\tau_c$ and $k_b$.}
	\label{fig:osc_hist}
\end{figure}

Table \ref{tab:sweep_results} summarizes the outcomes of all parameter combinations. Most solutions arrive at a stable fixed point. Only 0.2~\% of the solutions oscillate. Figure \ref{fig:osc_hist} shows the distribution of oscillatory solutions for each parameter range, with a dashed red line at each parameter's reference value. For all parameters, oscillatory solutions are much more likely with higher values. No oscillations occur in the lower end of the ranges for $a$, $C_w$, $k_b$, and $\tau_c$. Oscillation is particularly dependent on higher values of $k_b$ and $\tau_c$, where there is a complete lack of oscillatory solutions at and below the reference values. 

\begin{figure}[ht]
	\centering
	\includegraphics[width=\textwidth]{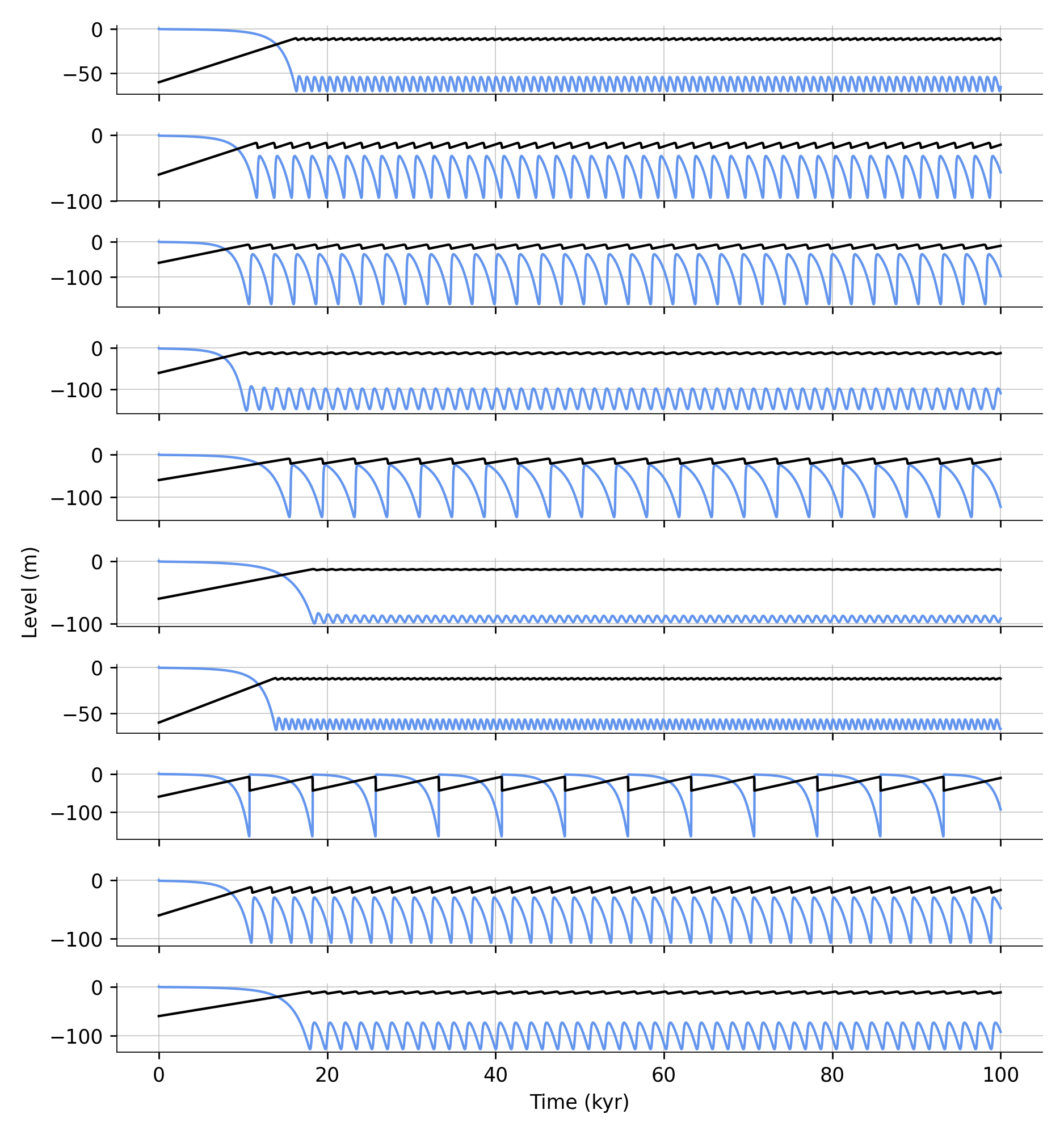}
	\caption{Ten oscillatory model solutions, with parameter sets chosen randomly from the group of solutions known to oscillate. The blue line shows $z_m$ and the black line shows $z_s$. The ocean level $z_o$ is omitted for clarity.}
	\label{fig:osc_ex}
\end{figure}

Figure \ref{fig:osc_ex} shows ten integrations with parameter combinations chosen randomly from the oscillatory solutions. Only the Mediterranean level (blue) and sill level (black) are shown, for clarity. The amplitudes of the oscillations here are all significantly less than in prior modeling, where MSL consistently reached -400~m and lower. The frequencies are also much higher. The oscillatory solutions tend to have many more than the roughly 16 cycles shown in previous work over 100~kyr. With higher values of $\tau_c$ and $a$, the system increasingly resembles a relaxation oscillator. The third simulation from the bottom of Figure \ref{fig:osc_ex} is a good example. Hydrological energy builds up as $z_m$ slowly drops, culminating in extremely rapid erosion of the sill and reflooding of the Mediterranean. Once the critical shear stress is exceeded, erosion grows exponentially. For higher values of $\tau_c$ and $a$, this transition from no erosion to very high erosion occurs extremely rapidly.

\section*{REVISITING CHANNEL SLOPE}

An important component of the model is the slope of the water surface in the channel connecting the Atlantic and Mediterranean. This slope controls erosion on the sill through equations (\ref{eq:zsdot}) and (\ref{eq:tau}). It also controls discharge into the Mediterranean through equations (\ref{eq:Q}) and (\ref{eq:V}). In this section, I analyze the choice of equation (\ref{eq:S}) for the channel slope and its consequences for model oscillation.

\subsection*{The Slope Equation}

\textcite{garcia-castellanos_messinian_2011} use equation (\ref{eq:S}) to express the average slope between the Atlantic and Mediterranean. This is a simple and clear expression. However, to calculate this average slope, they choose a constant horizontal distance $L$ of 100~km. It is not clear why the horizontal distance between the Atlantic and the shore of the Mediterranean would be constant as MSL varies over hundreds of meters in the model.

\begin{figure}[ht]
	\centering
	\includegraphics[width=0.7\textwidth]{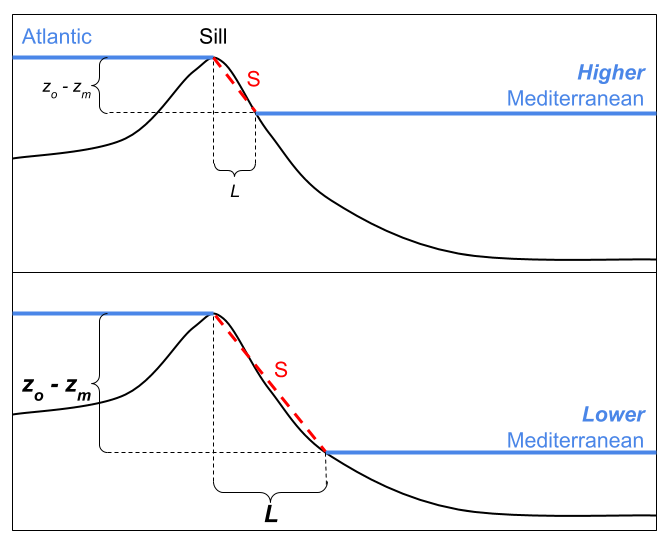}
	\caption{Schematic showing a cross-section from the Atlantic to the Mediterranean, through the sill. The top panel shows higher Mediterranean sea level and the bottom panel shows lower Mediterranean sea level. Both $z_o - z_m$ and $L$ would change as MSL changes. The average channel slope $S$ would be $(z_o - z_m)/L$, with variable $L$.}
	\label{fig:schematic}
\end{figure}

As the Mediterranean level rises and falls, the horizontal distance between the sill and the sea would change. The change in this horizontal distance would be governed by the bathymetry of the Mediterranean near the sill. For example, if MSL drops 100~m and the average slope of the newly exposed terrain is 1~\%, the horizontal length of the channel has increased 10~km. Figure \ref{fig:schematic} shows a schematic of this relationship between MSL, the channel length $L$, and the average channel slope $S$. More realistically, $L$ and $z_m$ would covary and the slope would be nearly constant for small changes in $z_m$.

\begin{figure}[ht]
	\centering
	\includegraphics[width=\textwidth]{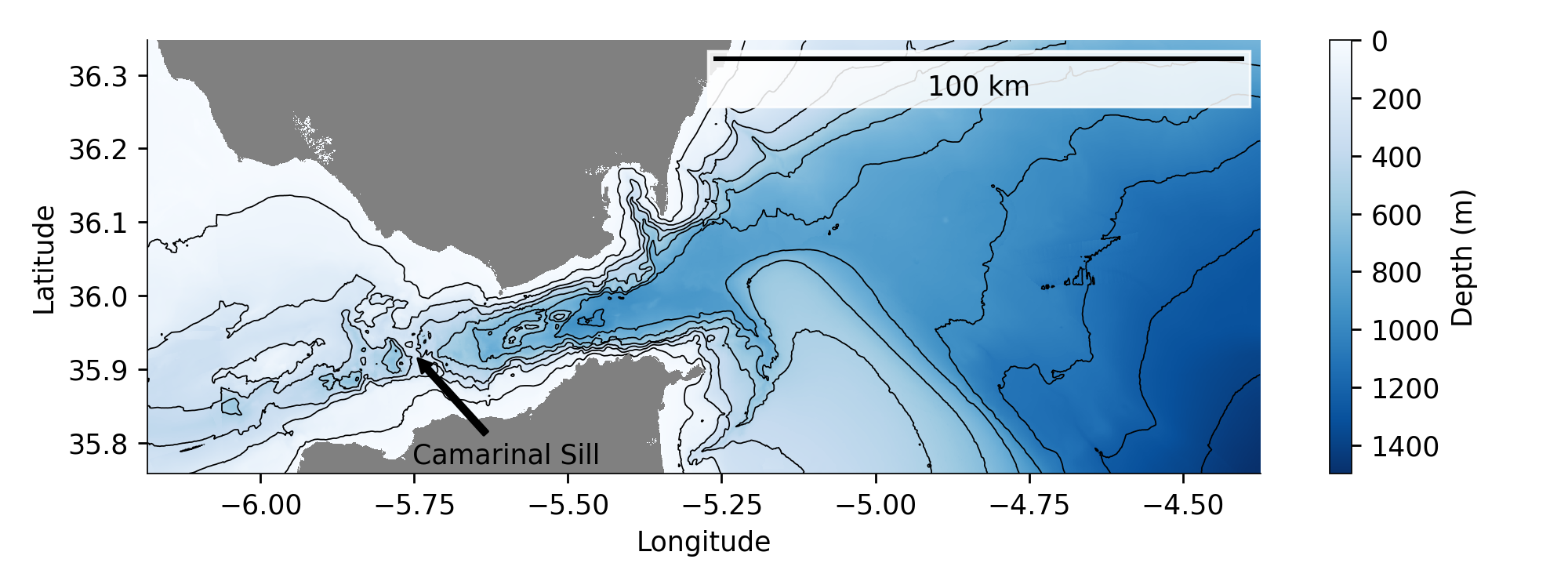}
	\caption{Bathymetry of the Strait of Gibraltar and western Mediterranean basin \parencite{emodnet}. Contours are shown at 150~m intervals. Grey area represents land.}
	\label{fig:bathymetry}
\end{figure}

To calculate the true average slope along the channel, detailed bathymetry would be required. We do not know the exact bathymetry of the Mediterranean during the MSC, but we can consider modern data for intuition. Figure \ref{fig:bathymetry} shows modern bathymetry at the Strait of Gibraltar and into the western Mediterranean basin. If modern bathymetry is any guide, there is little reason to expect a constant channel length of 100~km as MSL varies over hundreds of meters. In fact, the average slope would probably decrease when MSL drops, not increase, because deeper parts of the strait and nearby basin are flatter.

Further, the nonuniform slope through the modern strait raises doubts about whether the average slope is most applicable when MSL varies hundreds of meters. Although the average slope is simple and convenient, the local slope near the sill may be different than the average slope when MSL drops significantly. We do not know the shape of the sill during the Messinian, but we can consider the modern configuration again as an example. If MSL dropped below the Camarinal Sill today, erosion on the sill would likely be controlled by the higher local slope, at least until erosion significantly modified the sill's profile.

\begin{figure}[ht]
	\centering
	\includegraphics[width=0.7\textwidth]{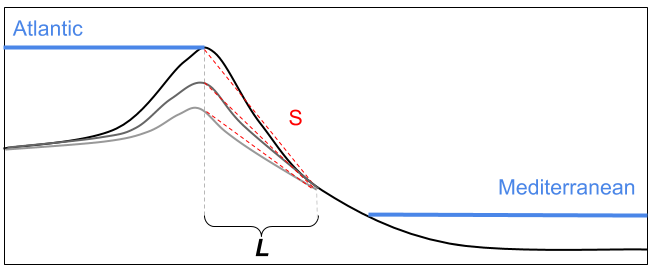}
	\caption{Schematic showing how the average slope might be computed if it depended on a fixed section near the sill and the sill height.}
	\label{fig:schematic2}
\end{figure}

In this scenario, where erosion occurs primarily over steeper terrain near the sill, the slope could theoretically be computed over a fixed length $L$ and depend on the height of the sill as it rises and falls. This conception is illustrated in Figure \ref{fig:schematic2}. However, this representation would require an equation for the slope that depends on sill height, $z_s$, and equation (\ref{eq:S}) does not depend on the sill height in any way. It depends only on $z_o$ and $z_m$, with constant $L$. Therefore, equation (\ref{eq:S}) does not represent an average local slope on the sill being slowly modified by uplift.

To summarize, if equation (\ref{eq:S}) is meant to represent the average slope between the Atlantic and Mediterranean sea levels, it should account for changing channel length $L$ instead of assuming $L$ is constant. There is no reason to expect Mediterranean sea level to rise and fall hundreds of meters without any lateral movement of the shoreline. If instead, equation (\ref{eq:S}) is meant to represent a more local slope near the sill as it is uplifted and eroded, it fails to capture this process because it does not depend on the sill height.

\subsection*{Implications for Oscillation}

Whichever vision of the average slope (Figure \ref{fig:schematic} or \ref{fig:schematic2}) is preferred, equation (\ref{eq:S}) is not sufficiently representative. This is an important concern, as the choice of equation (\ref{eq:S}) is critical to the model's capacity to oscillate. For oscillation to occur, there must be feedback that increases erosion when the Mediterranean level drops. In this model, a decrease in $z_m$ causes an increase in the slope $S$, strengthening erosion and opening the channel to flood the Mediterranean and raise $z_m$ again. The dependence of $S$ on $z_m$ creates the necessary feedback. Without this dependence, the feedback from $z_m$ to $\dot{z}_s$ is broken.

Crucially, it is the proposed form of equation (\ref{eq:S}), with average slope calculated using fixed $L$, that introduces the dependence of $S$ on $z_m$. Because changes in $z_o$ are much smaller than changes in $z_m$ ($<1$~\%), the slope in equation (\ref{eq:S}) is approximately a linear function of $z_m$. When $z_m$ drops, $S$ increases proportionally. However, as discussed above, the mean slope would probably be constant or decrease, not increase, if the change in $L$ and basin bathymetry are accounted for. Alternatively, if the local slope on the sill is more important than the mean slope, $S$ would not respond to the value of $z_m$ at all. In both cases, a more realistic representation of the slope is likely to render limit cycles impossible in this simple model because it would remove the proportional relationship between $S$ and $z_m$. When MSL drops, there would not be increased erosion on the sill to deepen the channel, refill the Mediterranean, and generate limit cycles.

\section*{DISCUSSION \& CONCLUSION}

In the Model Setup section, I introduced a simplified and consolidated version of a prior model used to explain the evaporite cycles of the first stage of the MSC \parencite{garcia-castellanos_messinian_2011}. The model is comprised of two ODEs with two supporting expressions and no need for additional data. It can be solved by any standard ODE integration method. In the Simulations section, I integrated the model for the same parameters used in the main results of prior work \parencite{garcia-castellanos_messinian_2011}, finding a stable solution without oscillation (Figure \ref{fig:refsim}). Then, using wide ranges of values for six key parameters, I checked all possible combinations of these values for oscillatory model solutions. Of the 64 million unique combinations checked, only 0.2~\% exhibit oscillation and these combinations are strongly skewed toward higher parameter values in each case. The simulations that do produce oscillations generally have amplitudes and frequencies quite different than those in prior work. It appears that limit cycles in the proposed model are not a robust result.

The discrepancy between the results here and the results of prior work is not readily explained. The only substantive change to the model equations used here is the treatment of the Mediterranean as a single basin. The prior model divides the eastern and western basins when $z_m$ falls below -430~m. There is, however, no obvious reason that this division would be necessary for oscillation. Similarly, although the Mediterranean area and ocean level are implemented with an explicit expression here instead of interpolation, this should not change the behavior of the model. Finally, it is always possible that different numerical implementations or programming mistakes may have influenced the results. Prior work implements a customized ``explicit finite-difference, time-iterative technique" to integrate the model, although this appears to simply be Euler's method \parencite{asalted}. For the present study, considerable effort was directed toward preventing numerical or programming problems. The simplified model was implemented in different programming languages and with distinct integration algorithms but yielded identical results for all tested parameter combinations. Additionally, both implementations rely on validated, publicly available integration codes \parencite{scipy, libode}.

In the Revisiting Channel Slope section, I explained how equation (\ref{eq:S}) is not physically representative of an average slope between the Atlantic and Mediterranean. To calculate the average channel slope, it assumes the horizontal position of the Mediterranean shoreline is fixed, even as the Mediterranean sea level varies tens or hundreds of meters. It is also not representative of the scenario where erosion is controlled by the local slope on the sill as the sill is uplifted and eroded. This is simply because equation (\ref{eq:S}) is completely independent of the sill height $z_s$. In either case, the channel slope should not significantly increase when $z_m$ decreases. The rate of erosion on the sill need not depend on Mediterranean sea level because the shore is downstream of the erosion process. This is a critical problem because the unrepresentative dependence of $S$ on $z_m$ introduced by equation (\ref{eq:S}) is required to produce oscillations in this model.

Future work could develop an improved expression for the model slope. However, this will likely be a nontrivial task. A correct formulation must handle the condition where $z_m>z_s$ and $z_o>z_s$. This is when the ocean is connected to the Mediterranean, as it is today with an open strait. When this happens, erosion and flow on the sill are poorly represented by idealized equations for stream dynamics. Proper treatment may have to dynamically account for periods when the sill is more like the ocean floor, instead of the bed of a stream. Further, as \textcite{coulson_role_2019} show, future work must also account for the gravitational and isostatic effects of the changing water loads.

In conclusion, the proposed model only rarely exhibits limit cycles, and limit cycles require parameter values considerably higher than those originally reported. However, the proposed model includes an unphysical representation of the channel slope. This is important because the proposed form of the channel slope is crucial to the model's capacity to oscillate. Properly treating the channel slope would make limit cycles unlikely for any parameter combinations and prior results should be reconsidered \parencite{garcia-castellanos_messinian_2011, coulson_role_2019}. A model with physically realistic slope equations might reveal whether sea-level limit cycles played a role in the mysterious first stage of the MSC, or if it must be explained by other mechanisms.

\subsubsection*{Acknowledgments}
I would like to thank two anonymous reviewers for very helpful feedback regarding the full scope and context of the MSC. I would also like to thank Jerry Mitrovica, Daniel Garcia-Castellanos, and Robin Wordsworth for their open, supportive, and objective discussions about this work. The primary computations in this paper were run on the FASRC Cannon cluster supported by the FAS Division of Science Research Computing Group at Harvard University. Integration methods from libode \parencite{libode} and SciPy \parencite{scipy} were used. Plots were created using the Matplotlib library \parencite{matplotlib}. All code used for the analysis and plots in this paper is permanently archived in Zenodo \parencite{msc_doi} and available with documentation at \url{github.com/markmbaum/messinian-salinity-crisis}.

\clearpage

\printbibliography

\clearpage

\begin{table}[ht]
  \centering
	\begin{tabular}{|c l|}
	  \hline
	  Parameter & Fit Value \\
	  \hline
	  $c_1$ & $2.068 \times 10^{12}$~m$^2$ \\
	  $\alpha_1$ & 2754~m \\
	  $c_2$ & $4.035 \times 10^{11}$~m$^2$ \\
	  $\alpha_2$ & 127.5~m \\
	  \hline
	\end{tabular}
  \caption{Parameter values for equation (\ref{eq:Am}), the surface area of the Mediterranean, and equation (\ref{eq:zo}), the ocean height.}
  \label{tab:Am}
\end{table}

\begin{table}
  \centering
  \begin{tabular}{|c c c c|}
    \hline
  	Parameter & Reference Value & Tested Range & Unit \\[1ex]
  	\hline
  	$k_b$ & 8$\times 10^{-6}$ & 1$\times 10^{-8}$ - 1$\times 10^{-4}$ & m/yr Pa$^\textrm{a}$ \\
  	$\tau_c$ & 50 & 25 - 100 & Pa \\
  	$C_w$ & 6 & 0.5 - 10 & - \\
  	$U$ & 4.9 & 0.49 - 4.9 & mm/yr \\
  	$a$ & 1.5 & 1 - 2 & - \\
  	$L$ & 100 & 50 - 150 & km \\
  	\hline
  	$n$ & 0.5 &  & s/m$^{1/3}$ \\
  	$P$ & 0.6 &  & m/yr \\
  	$E$ & 1.2 &  & m/yr \\
  	$R$ & 16500 &  & m$^3$/s \\
  	\hline
  \end{tabular}
  \caption{Reference values and tested ranges for each of the physical parameters in the model. Reference values are those used in the primary results of prior modeling \parencite{garcia-castellanos_messinian_2011}. For parameters with tested ranges, the model was integrated with all combinations of the values for each range. Where no tested range is shown, the parameter is left at the reference value for all simulations.}
  \label{tab:param}
\end{table}

\begin{table}
  \centering
  \begin{tabular}{|c l|}
    \hline
    Simulation Outcome & Percent\\
    \hline
  	disconnection & 20.9 \% \\
  	oscillation & 0.2 \% \\
    fixed point & 78.9 \% \\
    \hline
  \end{tabular}
  \caption{Outcomes of the 64 million simulations. ``Disconnection" refers to solutions where $z_s > z_o$, the Strait of Gibraltar closes, and the Mediterranean becomes completely disconnected from the Atlantic. ``Fixed point" refers to solutions that arrive at stable fixed points with an open channel between the Mediterranean and Atlantic, as in Figure \ref{fig:refsim}.}
  \label{tab:sweep_results}
\end{table}

\end{document}